# Application Component Placement in NFV-based Hybrid Cloud/Fog Systems

Carla Mouradian, Somayeh Kianpisheh, Roch H. Glitho
CIISE, Concordia University, Montreal, QC, Canada
{ca_moura, s_kianpi, glitho}@encs.concordia.ca

*Abstract*— Applications are sets of interacting components that can be executed in sequence, in parallel, or by using more complex constructs such as selections and loops. They can, therefore, be modeled as structured graphs with sub-structures consisting of these constructs. Fog computing can reduce the latency induced by distant clouds by enabling the deployment of some components at the edge of the network (i.e., closer to end-devices) while keeping others in the cloud. Network Functions Virtualization (NFV) decouples software from hardware and enables an agile deployment of network services and applications as Virtual Network Functions (VNFs). In NFV settings, efficient placement algorithms are required to map the structured graphs representing the VNF Forwarding Graphs (VNF-FGs) onto the infrastructure of the hybrid cloud/fog system. Only deterministic graphs with sequence and parallel sub-structures have been considered thus to date. However, several real-life applications do require non-deterministic graphs with sub-structures as selections and loops. This paper focuses on application component placement in NFV-based hybrid cloud/fog systems, with the assumption that the graph representing the application is non-deterministic. The objective is to minimize an aggregated weighted function of makespan and cost. The problem is modeled as an Integer Linear Programming (ILP) and evaluated over small-scale scenarios using the CPLEX optimization tool.

## I. INTRODUCTION

Many service providers use cloud computing to deploy their applications as a way to reduce cost whilst exploiting the elasticity feature provided by the cloud. However, the wide area network used to connect the cloud to the end-users might cause high latency, which may not be tolerable for some applications. On the other hand, fog computing, a concept introduced by CISCO in 2012, provides an intermediate layer between end-users and the cloud which allows the deployment of some of the application components in the fog at the edge while keeping some others in the cloud, thereby reducing latency [1].

Applications can be implemented in cloud/fog systems as a set of interacting components. Reference [2] discusses the example of a simple IoT application that detects fire and dispatches robots to fight the fire. The application consists of the following interacting components: a fire detector, fire-fighting strategies, and a robot dispatcher. Some applications could be more complex and involve more interacting components. An example is an IoT application for autonomous driving that enables cars to detect, track, and recognize objects such as cars and pedestrians, and to take the required actions. Fig. 1 shows a structured graph representation with the following sub-structures: sequence, selection, and loop. It should be noted that the selection sub-structure introduces non-determinism in the execution. For instance, in the case of immediate emergency situations, the Collision Avoidance component makes the decision to either "change the lane" or "perform emergency brake". The reader should note that probabilities could be assigned to these two alternatives.

Network Functions Virtualization (NFV) is an emerging technology that employs virtualization as a key technology. It aims at decoupling network functions from the underlying proprietary hardware and running them as software instances on general purpose hardware [3]. The network functions can be implemented as Virtualized Network Functions (VNFs). This paper assumes that an application's components are implemented as VNFs. The reader should note that beyond low-level network services, application components can also be implemented as VNFs (e.g., [4]). The structured graphs representing the applications are therefore VNF Forwarding Graphs (VNF-FG) (i.e., sets of VNFs chained in specific orders).

VNF-FG embedding (mapping of VNF-FGs onto NFV Infrastructure (NFVI)) is very challenging and has attracted much research interest [5]. However, to the best of our knowledge, no researcher has considered non-deterministic VNF-FG, although the implementation of real-life applications such as the autonomous driving mentioned above can lead to such graphs. Furthermore, most researchers focus on the traditional homogenous cloud as the setting for NFVI. Few works tackle the heterogeneity inherent in hybrid cloud/fog systems when it comes to the mapping of application components to a hybrid cloud/fog infrastructure.

This paper focuses on application component placement in NFV-based hybrid cloud/fog systems and tackles the challenges of heterogeneity and non-deterministic VNF-FG embedding. Heterogeneity is addressed by considering makespan (an important Quality of Service (QoS) criterion) in addition to cost (a budget that the application provider should pay for consuming resources) when it comes to optimization. Indeed, cost minimization encourages cloud usage while makespan minimization encourages fog usage. A compromise is required for the appropriate placement decision. The problem is modeled as an Integer Linear Programming (ILP) problem and evaluated over small-scale scenarios using the CPLEX optimization tool. We tackle the non-deterministic VNF-FG by assigning probabilities to selection sub-structures and mean numbers of iterations to loop sub-structures. These probabilities can be obtained through prediction models. The rest of the paper is organized as follows: Section II discusses the related work. The system model and problem formulation are explained in Section III, followed by the evaluation results in Section IV. We conclude the paper and outline likely future work in Section V.

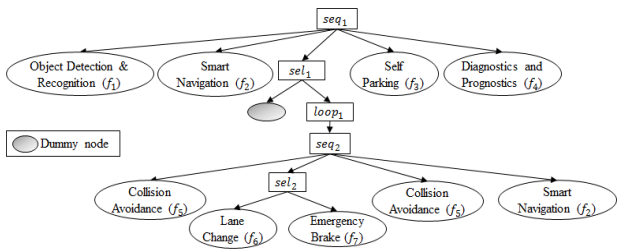

Fig. 1. A Structured Graph for Autonomous Driving Application

## II. RELATED WORKS

To the best of our knowledge, no researcher has addressed the problem of placing application components as VNFs in hybrid cloud/fog system. In this section, we first discuss the solutions proposed to date on application component placement in hybrid cloud/fog systems, even though these components are not placed as VNFs. We then discuss the research on VNF-FG embedding, although none of these efforts focuses on hybrid cloud/fog systems.

To solve the problem of application component placement in heterogeneous hybrid cloud/fog systems, authors in [2] propose an architecture for hybrid cloud/fog system. However, they do not provide any placement algorithm. In [6], researchers propose an algorithm that distributes the workload over the hybrid cloud/fog system considering the response time minimization and the throughput maximization. Power consumption minimization in hybrid cloud/fog systems was considered in [7]. Though these works do tackle the problem of component placement in heterogeneous hybrid cloud/fog systems, they do not take into account the cost imposed by infrastructure usage, another important criterion for component placement decision and one that conflicts with the time. In this paper, the two criteria of time and cost are considered. According to the weight of priority given to each criterion, the individual usage of cloud/fog or of a combination of both is decided.

The problem of VNF-FGs embedding in NFV and cloud networks has been studied widely over the last few years. Minimizing the infrastructure usage/cost aspects were investigated in [8] and [9]. The authors in [8] model the problem of placing a batch of VNF-FGs using ILP which minimizes the infrastructure utilization. Embedding VNF-FGs using a minimum number of VNF instances whilst meeting the end-to-end delay requirement was studied in [9]. The authors in [10] and [11] maximize the provider's revenue. In [10], the problem is solved by proposing online and offline methods. In the offline method, all requests are known in advance. The online method uses a prediction of future VNFs and their requirements. In [11], the VNF-FG embedding problem is modeled as a weighted graph matching problem and an eigendecomposition-based approach is proposed to solve it. All the above-mentioned works take as their input a set of deterministic VNF-FGs, where the sequence of VNFs is fixed in advance. However, in many cases, there are uncertainties associated with the VNFs' executions. These uncertainties imply that, for such non-deterministic input, from an initial state, there may be none, exactly one, or many possible transitions which are not considered in the literature. For instance, some VNFs run only when they are needed depending on the previous VNF's output, as in Fig.1, where the Collision Avoidance component is executed only if an emergency is detected by the Smart Navigation component.

## III. APPLICATION COMPONENT PLACEMENT PROBLEM DESCRIPTION

### A. Problem Description

The application component placement problem in NFV-based hybrid cloud/fog systems can be defined as follows: given a set of non-deterministic VNF-FGs representing applications, each VNF-FG is converted to a tree structure [12]. In the tree, leaf nodes represent application components, while middle nodes represent one of the sub-structures (e.g., loop, parallel, selection, sequence). The cost and the makespan are computed by aggregating the cost and the makespan of the nodes from bottom to top, according to the tree structure. The makespan and the cost of each component are computed first, and then these values are aggregated to calculate the makespan and the cost of the sub-structures in the tree. Finally, these values are composed according to the structure of the tree to generate the makespan and the cost of the VNF-FG.

In this problem, we aim at placing these non-deterministic VNF-FGs over a set of heterogeneous cloud and fog nodes, taking into consideration the short distance between IoT devices and fog nodes and the longer distance between IoT devices and the cloud nodes. The fog nodes have limited processing capabilities and are more expensive compared to cloud nodes; however, they provide lower latency. We assume that these components are implemented as VNFs, therefore the cloud and the fog infrastructures represent NFVI. We also consider that VNFs can be reused by more than one application. Our objective is to enable the embedding of VNF-FGs in a cloud and fog NFVIs at the lowest possible cost and makespan.

### B. System Model

We consider that the cloud and the fog domains are modeled as undirected graphs: $G^C = (N^C, E^C)$ and $G^F = (N^F, E^F)$, respectively. $N^C$ and $N^F$ are sets of physical cloud and fog nodes while $E^C$ and $E^F$ are sets of cloud and fog edges/links. We use $c_{n^C}, \gamma_{n^C}, c_{n^F}, \gamma_{n^F}$ to represent the capacity and the cost per unit of resource (e.g., CPU, memory, storage) of cloud nodes and fog nodes, respectively. $BW_{e^C}, \rho_{e^C}, BW_{e^F}, \rho_{e^F}$ are used to represent the bandwidth capacity and the transmission cost incurred by sending traffic along cloud edges and fog edges. $E^{CF}$ represents a set of edges between cloud and fog nodes, while $BW_{e^{CF}}$ and $\rho_{e^{CF}}$ denote the bandwidth capacity and the transmission cost of a cloud-fog edge per traffic unit.

$Req$ is a set of VNF-FG requests assigned to the system, and each VNF-FG request is indicated as $R \in Req$. $D$ denotes a set of IoT devices. The IoT devices, the traffic load, and the set of required VNF types for a request $R$ are indicated as $d^R, A^R$, and $V^R$, respectively. $T$ denotes the set of defined VNF types in the system, $f^t$ is a VNF of type $t \in T$ and can be shared by more than one request. $IP(f^t)$ represents the immediate predecessor of $f^t$, and the VNF edges are represented as $(f^t, IP(f^t))$. $BW_{d^R, n^C}$ and $BW_{d^R, n^F}$ are the bandwidth capacities between IoT devices and the cloud and fog nodes, respectively, and $\rho_{d^R, n^C}$ and $\rho_{d^R, n^F}$ are the transmission costs between IoT devices and cloud and fog nodes, respectively. Each VNF has a predefined license cost $\partial_{f^t}$, processing capacity $c_{f^t}$, and resource requirements $u_{f^t}$. The set of available instances for VNF type $t$ is represented as $I_{f^t}$.

Table I. Summary of key notations and variables

| Inputs | |
|---|---|
| $N^C, N^F$ | Set of cloud/fog nodes |
| $E^C, E^F, E^{CF}$ | Set of cloud/fog/cloud-fog edges |
| $Req$ | Set of requests assigned to the system |
| $R$ | Request for VNF-FG $R \in Req$ |
| $D$ | Set of IoT devices in the system |
| $A^R$ | Traffic units of request R |
| $d^R$ | Set of IoT devices used by request $R$ |
| $V^R$ | Set of required VNFs for request $R$ |
| $T$ | Set of VNF types |
| $f^t$ | VNF of type $t \in T$ |
| $IP(f^t)$ | Immediate predecessor of $f^t$ |
| $fst^R$ | First VNF in a VNF-FG request $R$ |
| $\partial_{f^t}, c_{f^t}, u_{f^t}$ | License cost, capacity, requirements of VNF type $t$ |
| $I_{f^t}$ | Set of instances for VNF type $t$ |
| $S_i$ | Basic sub-structure or VNF |
| $it$ | Expected number of iterations of a loop |
| $p_i$ | Probability of selection |
| $c_{n^C}, c_{n^F}$ | Capacity of cloud/fog nodes (per resource unit) |
| $\gamma_{n^C}, \gamma_{n^F}$ | Cost of cloud/fog nodes per resource unit |
| $BW_{e^C}, BW_{e^F}, BW_{e^{CF}}$, $BW_{d^R, n^C}, BW_{d^R, n^F}$ | Bandwidth capacity of cloud/fog/cloud-fog/IoT-cloud/IoT-fog edges per traffic unit |
| $\rho_{e^C}, \rho_{e^F}, \rho_{e^{CF}}$, $\rho_{d^R, n^C}, \rho_{d^R, n^F}$ | Transmission cost of cloud/fog/cloud-fog/IoT-cloud/IoT-fog edges per traffic unit |
| $D_{f^t}^{n^C}, D_{f^t}^{n^F}$ | Processing delay of VNF type $t$ on cloud/fog nodes |
| $D^{e^C}, D^{e^F}, D^{e^{CF}}$, $D_{d^R}^{n^C}, D_{d^R}^{n^F}$ | Cloud/fog/cloud-fog/IoT-cloud/IoT-fog edges delays per traffic unit |
| $\mu$ | Maximum nodes/edges/VNF usage threshold |
| Variables | |
| $x_{i,f^t,n^C}, x_{i,f^t,n^F}$ | Binary variable, indicating if instance $i$ of VNF type $t$ is instantiated on cloud/fog node |
| $x_{i,f^t,n^C}^R, x_{i,f^t,n^F}^R$ | Binary variable, indicating if instance $i$ of VNF type $t$ instantiated on cloud/fog node is assigned to request $R$ |
| $y_{f^t, IP(f^t), e^C}^R$, $y_{f^t, IP(f^t), e^F}^R$, $y_{f^t, IP(f^t), e^{CF}}^R$ | Binary variable, indicating if VNF edge $(f^t, IP(f^t))$ mapped to cloud/fog/cloud-fog edges is assigned to request $R$ |
| $y_{n^C, d^R}^R, y_{n^F, d^R}^R$ | Binary variable indicating if a link is assigned between IoT devices and cloud/fog nodes |

We construct a tree for each VNF-FG, in which a composition of four sub-structures can be used: sequence, parallel, selection, and loop. $S_i$ represents a sub-structure or a VNF, $S_i \in \{seq, par, sel, loop\} \cup V^R$. $it$ is the expected number of iterations of a loop and is defined as: $it = \sum_{n=0}^{\infty} n \cdot q^n$ where $n$ is the number of iterations and $q$ is the probability of loop occurrence. Since $-1 \leq |q| \leq 1$ then the partial sum of this infinite series converges and is calculated as $\frac{q}{1-q}$. $p_i$ is the probability of selection of $S_i$. $p_i = 1$ for $seq, par, loop$.

The delays are represented as follows: $D_{f^t}^{n^C}$ and $D_{f^t}^{n^F}$ are the delays per traffic unit for processing VNF of type $t$ on cloud node $n^C \in N^C$ and fog node $n^F \in N^F$ respectively, while $D^{e^C}$, $D^{e^F}$, and $D^{e^{CF}}$ are the delays per traffic unit of cloud edge $e^C \in E^C$, fog edge $e^F \in E^F$, and cloud-fog edge $e^{CF} \in E^{CF}$, respectively. $D_{d^R}^{n^C}$ and $D_{d^R}^{n^F}$ are the delays per traffic unit between IoT devices and the cloud and fog nodes respectively.

### C. Problem Formulation

We formulate our problem as an ILP with the objective of minimizing the aggregated weighted function of cost and makespan. We define the following decision variables:

- $x_{i,f^t,n^C} \in \{0,1\}$ and $x_{i,f^t,n^F} \in \{0,1\}$ are equal to 1 if instance $i$ of VNF $f^t$ is mapped to cloud node $n^C \in N^C$ or fog node $n^F \in N^F$, respectively.
- $x_{i,f^t,n^C}^R \in \{0,1\}$ and $x_{i,f^t,n^F}^R \in \{0,1\}$ are equal to 1 if instance $i$ of VNF $f^t$ mapped to cloud node $n^C \in N^C$ or fog node $n^F \in N^F$ respectively, is assigned to request $R$.
- $y_{f^t, IP(f^t), e^C}^R \in \{0,1\}$, $y_{f^t, IP(f^t), e^F}^R \in \{0,1\}$, and $y_{f^t, IP(f^t), e^{CF}}^R \in \{0,1\}$ are equal to 1 if VNF edge $(f^t, IP(f^t))$ on a cloud edge $e^C \in E^C$, a fog edge $e^F \in E^F$, or a cloud-fog edge $e^{CF} \in E^{CF}$, respectively, are assigned to request $R$.
- $y_{d^R, n^C}^R \in \{0,1\}$ and $y_{d^R, n^F}^R \in \{0,1\}$ are equal to 1 if a link is assigned between an IoT device $d^R \in D$ and cloud node $n^C \in N^C$ or fog node $n^F \in N^F$, respectively.

Table I summarizes the ILP parameters and variables.

#### 1) Expected Cost Value

As stated above, a node in the tree of the input VNF-FG can be one of the four sub-structures or a VNF. The costs for the four sub-structures are given in Table II, while the cost for $S_i \in V^R$ is described below.

*a. Processing Cost:* the cost of resources assigned to all VNFs belonging to a VNF-FG request:

$$C_{proc}(f^t) = \sum_{n^C \in N^C} \sum_{i \in I_{f^t}} x_{i,f^t,n^C}^R \cdot \gamma_{n^C} \cdot u_{f^t} + \sum_{n^F \in N^F} \sum_{i \in I_{f^t}} x_{i,f^t,n^F}^R \cdot \gamma_{n^F} \cdot u_{f^t} \quad (1)$$

The total processing cost of a request is thus:
$$C_{proc}(R) = C_{proc}(root) \quad (2)$$

*b. Deployment Cost:* the cost of total software license costs for the deployment of VNF instances:

$$C_{dep}(R) = \sum_{n^C \in N^C} \sum_{f^t \in V^R} \sum_{i \in I_{f^t}} x_{i,f^t,n^C}^R \cdot \partial_{f^t} + \sum_{n^F \in N^F} \sum_{f^t \in V^R} \sum_{i \in I_{f^t}} x_{i,f^t,n^F}^R \cdot \partial_{f^t} \quad (3)$$

*c. Communication Cost:* the cost of assigned edges to all VNF edges in a request, including the communication cost between IoT devices and their corresponding VNFs on cloud and/or fog nodes:

$$C_{com}(f^t) = \sum_{e^C \in E^C} y_{f^t, IP(f^t), e^C}^R \cdot A^R \cdot \rho_{e^C} + \sum_{e^F \in E^F} y_{f^t, IP(f^t), e^F}^R \cdot A^R \cdot \rho_{e^F} + \sum_{e^{CF} \in E^{CF}} y_{f^t, IP(f^t), e^{CF}}^R \cdot A^R \cdot \rho_{e^{CF}} \quad (4)$$

The total communication cost of a request is:
$$C_{com}(R) = C_{com}(root) + \sum_{n^C \in N^C} \sum_{d^R \in D} y_{d^R, n^C}^R \cdot A^R \cdot \rho_{d^R, n^C} + \sum_{n^F \in N^F} \sum_{d^R \in D} y_{d^R, n^F}^R \cdot A^R \cdot \rho_{d^R, n^F} \quad (5)$$

The total expected cost for a VNF-FG request is then:
$$C(R) = C_{proc}(R) + C_{dep}(R) + C_{com}(R) \quad (6)$$

#### 2) Expected Makespan Value

The expected makespan value composed of the processing times and the communication times, for the four sub-structures is given in Table II. For $S_i \in V^R$, the processing time and the communication time are given below. It should be noted that the processing and communication time equations in Table II for parallel sub-structure are non-linear. However, they can be linearized by replacing them with Eq. (7-1) and (7-2) for processing time:

$$z = \max_{i=1,\ldots,n} M_{proc}(S_i) \quad (7\text{-}1)$$
$$z \geq M_{proc}(S_i) \quad \forall i = 1..n \quad (7\text{-}2)$$

a. **Processing Time:** the delay of processing the deployed VNFs belonging to a VNF-FG request.

Table II. The cost and the makespan estimation for $S_i \in \{seq, par, sel, loop\}$

| Sub-structures | Processing Cost $C_{proc}(S_i)$ | Communication Cost $C_{com}(S_i)$ | Processing Time $M_{proc}(S_i)$ | Communication Time $M_{com}(S_i)$ |
|---|---|---|---|---|
| $S_i$ is seq | $\sum_{i=1}^{n} C_{proc}(S_i)$ | $\sum_{i=1}^{n} C_{com}(S_i)$ | $\sum_{i=1}^{n} M_{proc}(S_i)$ | $\sum_{i=1}^{n} M_{com}(S_i)$ |
| $S_i$ is par | $\sum_{i=1}^{n} C_{proc}(S_i)$ | $\sum_{i=1}^{n} C_{com}(S_i)$ | $\max_{i=1,\ldots,n} M_{proc}(S_i)$ | $\max_{i=1,\ldots,n} M_{com}(S_i)$ |
| $S_i$ is sel | $\sum_{i=1}^{n} p_i \cdot C_{proc}(S_i)$ | $\sum_{i=1}^{n} p_i \cdot C_{com}(S_i)$ | $\sum_{i=1}^{n} p_i \cdot M_{proc}(S_i)$ | $\sum_{i=1}^{n} p_i \cdot M_{com}(S_i)$ |
| $S_i$ is loop | $it. \, C_{proc}(seq)$ | $it. \, C_{com}(S_i)$ | $it. \, M_{proc}(seq)$ | $it. \, M_{com}(seq)$ |

$$M_{proc}(f^t) = \sum_{n^C \in N^C} \sum_{i \in I_{f^t}} x^R_{i,f^t,n^C} \cdot A^R \cdot D^{n^C}_{f^t} + \sum_{n^F \in N^F} \sum_{i \in I_{f^t}} x^R_{i,f^t,n^F} \cdot A^R \cdot D^{n^F}_{f^t} \quad (8)$$

The total processing time of a VNF-FG request is then:
$$M_{proc}(R) = M_{proc}(root) \quad (9)$$

b. **Communication Time:** the communication time includes the communication of all VNF edges assigned in a request and the communication time between the IoT devices and their corresponding VNFs on cloud or fog nodes.

$$M_{com}(f^t) = \sum_{e^C \in E^C} y^R_{f^t,IP(f^t),e^C} \cdot A^R \cdot D^{e^C} + \sum_{e^F \in E^F} y^R_{f^t,IP(f^t),e^F} \cdot A^R \cdot D^{e^F} + \sum_{e^{CF} \in E^{CF}} y^R_{f^t,IP(f^t),e^{CF}} \cdot A^R \cdot D^{e^{CF}} \quad (10)$$

The total communication time of a VNF-FG request thus:
$$M_{com}(R) = M_{com}(root) + \sum_{n^C \in N^C} \sum_{d^R \in D} y^R_{d^R,n^C} \cdot A^R \cdot D^{n^C}_{d^R} + \sum_{n^F \in N^F} \sum_{d^R \in D} y^R_{d^R,n^F} \cdot A^R \cdot D^{n^F}_{d^R} \quad (11)$$

The expected makespan value of a VNF-FG request is then:
$$M(R) = M_{proc}(R) + M_{com}(R) \quad (12)$$

*3) Objective Function*

Our objective is to minimize the makespan and the cost of a set of VNF-FGs, as shown in Eq. (13).

$$Min \, (\alpha \sum_{\forall R \in Req} C(R) + (1-\alpha) \sum_{\forall R \in Req} M(R)) \quad (13)$$

where $\alpha$ is the weight of the objective functions that defines priorities among them, $1 \geq \alpha \geq 0$. E.g., $\alpha = 1$ motivates placement on cloud while $\alpha = 0$ motivates placement in fog.

*4) Constraints*

- **Nodes capacity constraint:** Eq. (14-1) and (14-2) ensure that the cloud and fog nodes are not overloaded.

$$\sum_{t \in T} \sum_{i \in I_{f^t}} u_{f^t} \cdot x^R_{i,f^t,n^C} \leq \mu_{n^C} \cdot c_{n^C} \quad \forall n^C \in N^C \quad (14\text{-}1)$$

$$\sum_{t \in T} \sum_{i \in I_{f^t}} u_{f^t} \cdot x^R_{i,f^t,n^F} \leq \mu_{n^F} \cdot c_{n^F} \quad \forall n^F \in N^F \quad (14\text{-}2)$$

- **Edges capacity constraint:** Eq. (15-1) to (15-5) ensure that the cloud, fog, cloud-fog, IoT-cloud, and IoT-fog edges assigned to the requests are not overloaded.

$$\sum_{\forall R \in Req} A^R \cdot y^R_{f^t,IP(f^t),e^C} \leq \mu_{e^C} \cdot BW_{e^C} \quad \forall e^C \in E^C \quad (15\text{-}1)$$

$$\sum_{\forall R \in Req} A^R \cdot y^R_{f^t,IP(f^t),e^F} \leq \mu_{e^F} \cdot BW_{e^F} \quad \forall e^F \in E^F \quad (15\text{-}2)$$

$$\sum_{\forall R \in Req} A^R \cdot y^R_{f^t,IP(f^t),e^{CF}} \leq \mu_{e^{CF}} \cdot BW_{e^{CF}} \quad \forall e^{CF} \in E^{CF} \quad (15\text{-}3)$$

$$\sum_{\forall R \in Req} A^R \cdot y^R_{d^R,n^C} \leq \mu_{d^R,n^C} \cdot BW_{d^R,n^C} \quad \forall d^R \in D, n^C \in N^C \quad (15\text{-}4)$$

$$\sum_{\forall R \in Req} A^R \cdot y^R_{d^R,n^F} \leq \mu_{d^R,n^F} \cdot BW_{d^R,n^F} \quad \forall d^R \in D, n^F \in N^F \quad (15\text{-}5)$$

- **Edge assignment constraints:** Eq. (16-1) and (16-2) ensure that an edge is assigned between a selected cloud or fog node for the first VNF of request $R$ and the IoT devices. We assume that the first VNF in the VNF-FG is the one that communicates with the IoT devices. Eq. (17), (18), and (19) ensure that an edge is assigned between VNFs for each request $R$.

$$x^R_{i,fst^R,n^C} = y^R_{n^C,d^R} \quad \forall n^C \in N^C, R \in Req, i \in I_{fst^R}, d^R \in D \quad (16\text{-}1)$$

$$x^R_{i,fst^R,n^F} = y^R_{n^F,d^R} \quad \forall n^F \in N^F, R \in Req, i \in I_{fst^R}, d^R \in D \quad (16\text{-}2)$$

$$x^R_{i,f^t,s} \cdot x^R_{j,IP(f^t),t} \leq y^R_{f^t,IP(f^t),e^C}$$
$$\forall s,t \in N^C, R \in Req, f^t \in V^R, i \in I_{f^t}, j \in I_{IP(f^t)} \quad (17)$$

$$x^R_{i,f^t,s} \cdot x^R_{j,IP(f^t),t} \leq y^R_{f^t,IP(f^t),e^F}$$
$$\forall s,t \in N^F, R \in Req, f^t \in V^R, i \in I_{f^t}, j \in I_{IP(f^t)} \quad (18)$$

$$x^R_{i,f^t,s} \cdot x^R_{j,IP(f^t),t} \leq y^R_{f^t,IP(f^t),e^{CF}}$$
$$\forall s \in N^C, t \in N^F, R \in Req, f^t \in V^R, i \in I_{f^t}, j \in I_{IP(f^t)} \quad (19)$$

Eq. (17), (18), and (19) are non-linear, however, they can be linearized by replacing them with linear equations. Eq. (17-1) (17-5) show how Eq. (17) is linearized. In a similar way Eq. (18) and (19) are linearized:

$$Q^R_{s,t,f^t,IP(f^t),i,j} = x^R_{i,f^t,s} \cdot x^R_{j,IP(f^t),t} \quad (17\text{-}1)$$
$$\forall s,t \in N^C, R \in Req, f^t \in V^R, i \in I_{f^t}, j \in I_{IP(f^t)}$$

$$Q^R_{s,t,f^t,IP(f^t),i,j} \leq y^R_{f^t,IP(f^t),e^C} \quad (17\text{-}2)$$
$$\forall s,t \in N^C, R \in Req, f^t \in V^R, i \in I_{f^t}, j \in I_{IP(f^t)}$$

$$Q^R_{s,t,f^t,IP(f^t),i,j} \leq x^R_{i,f^t,s} \quad (17\text{-}3)$$
$$\forall s,t \in N^C, R \in Req, f^t \in V^R, i \in I_{f^t}, j \in I_{IP(f^t)}$$

$$Q^R_{s,t,f^t,IP(f^t),i,j} \leq x^R_{j,IP(f^t),t} \quad (17\text{-}4)$$
$$\forall s,t \in N^C, R \in Req, f^t \in V^R, i \in I_{f^t}, j \in I_{IP(f^t)}$$

$$Q^R_{s,t,f^t,IP(f^t),i,j} \leq x^R_{i,f^t,s} + x^R_{j,IP(f^t),t} - 1 \quad (17\text{-}5)$$
$$\forall s,t \in N^C, R \in Req, f^t \in V^R, i \in I_{f^t}, j \in I_{IP(f^t)}$$

- **VNF capacity constraint:** Eq. (20-1) and (20-2) ensure that the VNFs are not overloaded.

$$\sum_{\forall R \in Req} A^R \cdot x^R_{i,f^t,n^C} \leq \mu_{f^t} \cdot c_{f^t} \quad \forall t \in T, i \in I_{f^t}, n^C \in N^C \quad (20\text{-}1)$$

$$\sum_{\forall R \in Req} A^R \cdot x^R_{i,f^t,n^F} \leq \mu_{f^t} \cdot c_{f^t} \quad \forall t \in T, i \in I_{f^t}, n^F \in N^F \quad (20\text{-}2)$$

Table III. Simulation Parameters

| Parameter | Value |
|---|---|
| Number of nodes: cloud, fog | 2,3 |
| Number of IoT devices $D$ | 5 |
| Bandwidth capacity: IoT-fog, fog-fog, fog-cloud, IoT-cloud, Cloud-cloud | 54Mbps, 100 Mbps, 10 Gbps, 10 Gbps, 100 Gbps |
| Edges delay (ms): IoT-fog, fog-fog, fog-cloud, IoT-cloud, Cloud-cloud | Rand [1,2], Rand [0.5,1.2], Rand [15, 35], Rand [15,35], 0.64 |
| Edges bandwidth cost ($/Gb): IoT-fog, fog-fog, fog-cloud, IoT-cloud, cloud-cloud | 1, 2 , 3, 4, 0.1 |
| Node capacity (vCPU): cloud, fog | 8, 3 |
| Nodes cost ($/vCPU): cloud, fog | 0.1, 6 |
| Number of requests (i.e., application) | [1-3] |
| Number of VNFs in app1, app2, app3 | [6, 6, 7] |
| VNF license cost ($) | 100 |
| VNF resource requirements (vCPU) | Rand [1-4] |
| VNF processing delay (ms): cloud, fog | 3.12, 0.03 |
| Maximum usage threshold | 1 |
| Traffic load (KB) | 80 |
| q, p, it | 0.25, 0.5, 0.33 |

- **VNF assignment constraints:** Eq. (21) ensures that only one instance of each required VNF type is assigned to request $R$. Eqs. (22-1) and (22-2) ensure that the assigned VNF instances are already deployed in the network.

$$\sum_{\forall n^C \in N^C} \sum_{\forall i \in I_{ft}} x_{i,f^t,n^C}^R + \sum_{\forall n^F \in N^F} \sum_{\forall i \in I_{ft}} x_{i,f^t,n^F}^R = 1 \quad \forall f^t \in V^R, R \in Req \quad (21)$$

$$x_{i,f^t,n^C}^R \leq x_{i,f^t,n^C} \quad \forall R \in Req, f^t \in V^R, i \in I_{ft}, n^C \in N^C \quad (22-1)$$

$$x_{i,f^t,n^F}^R \leq x_{i,f^t,n^F} \quad \forall R \in Req, f^t \in V^R, i \in I_{ft}, n^F \in N^F \quad (22-2)$$

- **VNF deployment constraint:** Eq. (23) ensures that at least one instance of each required VNF type is deployed.

$$\sum_{\forall n^C \in N^C} \sum_{\forall i \in I_{ft}} x_{i,f^t,n^C} + \sum_{\forall n^F \in N^F} \sum_{\forall i \in I_{ft}} x_{i,f^t,n^F} \geq 1 \quad \forall t \in T \quad (23)$$

IV. PERFORMANCE EVALUATION

In this section, we first present the simulation setup, then the evaluation scenario followed by the obtained results. The ILP model is implemented and solved in IBM CPLEX 12.08. All simulations were conducted on a single machine with dual 2X8-Core 2.50GHz Intel Xeon CPU E5-2450v2 and 40GB memory.

*A. Simulation Setup*

We consider network topology with 5 IoT devices, 2 cloud nodes, and 3 fog nodes; cloud and fog nodes are fully connected. The IoT devices do not process data, they either send it to cloud or fog nodes for processing and analysis. We assume the same traffic amount passing all VNFs. For cost and makespan values, we use the same range of values with the same decimal scaling to guarantee meaningful comparison between them. We used the same conditions and parameter settings of simulation scenarios in the literature (i.e., [6][13][14][15][16][17]) to obtain meaningful results. All simulation parameters are summarized in Table III. We assume 3 applications: (1) earthquake early warning application with 6 VNFs, (2) flood warning application with 6 VNFs, and (3) autonomous driving application with 7 VNFs (Fig. 1).

Table IV. Comparison of cost, makespan, aggregated weighted function of cost and makespan when using only cloud, only fog, and hybrid cloud/fog for the three application

| Apps | Cost ($) | | | Makespan (ms) | | | Aggregated weighted function of cost and makespan $\alpha = 0.5$ | | |
|---|---|---|---|---|---|---|---|---|---|
| | Cloud | Fog | Cloud/Fog | Cloud | Fog | Cloud/Fog | Cloud | Fog | Cloud/Fog |
| App1 | 602.0 | 760 | 641.5 | 180 | 4.80 | 41.1 | 391.0 | 382.4 | 341.32 |
| App2 | 602.1 | 768 | 649.5 | 180 | 4.81 | 76.4 | 391.0 | 386.4 | 362.99 |
| App3 | 700.6 | 753 | 708.9 | 180 | 5.85 | 40.6 | 440.3 | 379.7 | 374.78 |

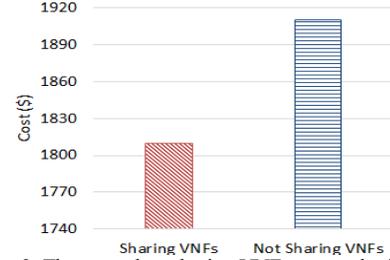

Fig. 2. The cost when sharing VNFs vs not sharing

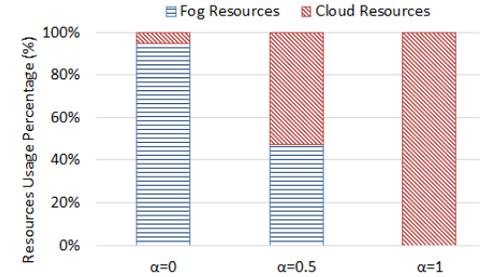

Fig. 3. Resources usage percentage when varying $\alpha$

Applications 1 and 2 share VNF of the same type, i.e., Historical Storage. We assume the VMs hosting the VNFs range from tiny to large size OpenStack VM, where large-size VM indicates a computationally intensive VNF. For application 3 (Fig. 1), we consider the probability of collision risk $q = 0.25$ [18], and hence $it=0.33$. We give equal probabilities for selection; $p=0.5$.

*B. Evaluation Scenario*

We implemented multiple scenarios with a different number of applications (i.e., requests). We considered scenarios with only fog domain, only cloud domain, and hybrid cloud/fog systems to evaluate their impact on the component placement problem. In addition, for application 1 and 2, we implemented the case where they do not share VNFs to obtain some insights on the gain of reusing VNFs by more than one application. Also, we varied the value of $\alpha$ to show the percentages of resources usage in cloud and fog. Finally, we compared the cost and the makespan to the one obtained based on a random placement of VNFs that still satisfies the nodes and edges capacities.

*C. Evaluation Results*

We solved the problem over the scenarios described in the previous section. Table IV shows the cost, the makespan, and the objective function for each application. We can see that the best results for the aggregated weighted function of cost and makespan are obtained when the components are placed on a hybrid cloud/fog system. As can also be observed, when only using cloud the cost is minimized, however, the makespan is increased. Similarly, when only fog is used, the makespan is minimized but the cost is increased. This is because fog provides lower latency than cloud due to its proximity to end-devices, but the resources in fog are more expensive. It should be noted that when using only fog, it is ensured that the total resources required by VNFs do not exceed fog domain capacity. Fig. 2 shows the cost when two applications using the same type of VNFs share vs. when they do not share. As expected, the cost is much lower when sharing is supported. This shows the gain of reusing VNFs by more than one application.

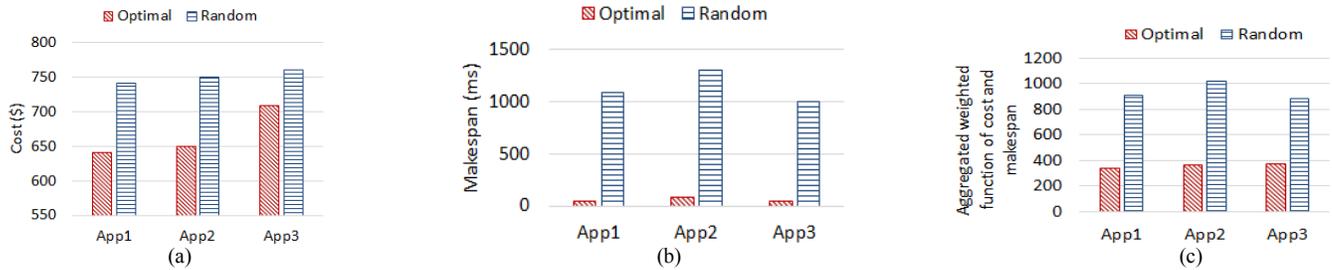

Fig. 4. Comparison between solutions obtained from optimal placement vs. solutions obtained from random placement
a) Cost, b) Makespan, c) Aggregated weighted function of cost and makespan $\alpha = 0.5$

Fig. 3 shows the percentages of resource usage in cloud and fog domains when varying $\alpha$. As expected, when $\alpha = 0$, most of the resources used are those in the fog domain since higher priority is given to the makespan, some resources are still used in the cloud because of the fog nodes capacity constraints. As $\alpha$ increases, more resources in the cloud domain are used and when $\alpha = 1$, all the resources are used from cloud domain. We also compared our results to the one obtained based on a random placement. We can see in Fig. 4(a, b, c) that there are notable differences between our model and a random placement with the latter leading to higher cost, makespan and aggregated function of both. Overall, these observations highlight the importance of considering a strategic mapping of application components to hybrid cloud/fog infrastructure for the efficient management of resources.

## V. CONCLUSION

We have addressed the application component placement problem in NFV-based hybrid cloud/fog systems by tackling two main challenges; the heterogeneity of cloud and fog systems and non-deterministic VNF-FGs. The heterogeneity is tackled by considering both makespan and cost while the non-determinism challenge is addressed by assigning probabilities and a mean number of iterations for selection and loop sub-structures. We modeled the problem as an ILP formulation and evaluated it considering different scenarios. We also compared its outcome to that of a random placement. Our results show that random placement results in higher cost and time than the optimal solution. This indicates the need for an efficient component placement algorithm. In the future, we plan to extend this work by designing efficient component placement algorithm and evaluate it considering large-scale scenarios.

ACKNOWLEDGMENT

This work is partially funded by the Canada Research Chair Program and by the Canadian Natural Sciences and Engineering Research Council (NSERC) through a Discovery Grant.